\documentclass{article}
\usepackage{graphicx} 
\usepackage[margin=2cm]{geometry}
\usepackage{siunitx}
\usepackage{amsmath}
\usepackage[rightcaption]{sidecap}
\sidecaptionvpos{figure}{c}

\usepackage[
backend=biber,
style=nature,
sorting=none
]{biblatex}
\makeatletter
\renewcommand{\maketitle}{\bgroup\setlength{\parindent}{0pt}
\begin{flushleft}
  \textbf{\huge\@title}

  \Large\@author

  \vspace{1ex}

  \large\@date
\end{flushleft}\egroup
}
\makeatother

\title{\begin{flushleft}Approximations of MINFLUX Localization Precision with Background\end{flushleft}}
\date{}
\author{Zach Marin$^{1,2}$, Jonas Ries$^{1,2,3,4}$}
\date{
$^1 $Max Perutz Labs, Vienna Biocenter Campus (VBC), Vienna, Austria\\
$^2 $University of Vienna, Center for Molecular Biology, Department of Structural and \\
\hskip1ex Computational Biology, Vienna, Austria\\
$^3 $European Molecular Biology Laboratory, Cell Biology and Biophysics, Heidelberg, Germany\\
$^4 $University of Vienna, Faculty of Physics, Vienna, Austria\\
Correspondence: jonas.ries@univie.ac.at}

\begin{document}

\maketitle

\section*{Abstract}

MINFLUX \cite{balzarotti_nanometer_2017} is an emerging super-resolution technology that measures the position of single fluorophores with nanometer precision using fewer photons than any other fluorescence imaging or tracking technique. Here, we derive simple and instructive analytical equations for MINFLUX localization precision with a special focus on background photons. A fluorescence background, either arising from an imperfect zero of the MINFLUX excitation point spread function (PSF) or from auto- or out-of-focus fluorescence, ultimately limits the resolution achievable with MINFLUX. Building on previous work \cite{balzarotti_nanometer_2017, eilers_minflux_2018, masullo_common_2022}, we try to improve our understanding of the influence of background, especially when it is unknown, through a new set of expressions for the localization precision, based on an explicit background term instead of signal-to-background ratio. We use these equations to generate an intuitive understanding of how fluorescence background affects MINFLUX measurements and illustrate that:
\begin{itemize}
    \item The precision of an emitter position estimate depends on the gradient of its excitation profile.
    \item Knowledge of the fluorescence background, obtained during post-processing of MINFLUX traces or through separate measurements, provides a better localization precision than in the case of unknown background.
    \item In diffraction-limited systems, localization with a PSF that features a near-zero minimum outperforms localization with a maximum.
\end{itemize}

We also present an analytical expression for the localization precision in orbital tracking \cite{enderlein_tracking_2000}, which we use for comparison to MINFLUX. Non-trivial derivations presented in this paper can be found in the accompanying Mathematica notebook.

\section{Deriving the Cram\'{e}r-Rao bound for $K$ measurements of an emitter in the presence of background in 1D} \label{sec:1dcrb}

There is an emitter sitting on a line. We want to know its position $x_0$, so we probe it with an excitation PSF of shape $f(x)$ positioned at $x_i$. We assume the pinhole is large, so we always collect all of the emission signal. The intensity measured from an emitter positioned at $x_0$ can be expressed as
\begin{align} \label{eq:Igen}
    I(x_0-x_i) = I_0(f(x_0-x_i) + b),
\end{align}
where $I(x_0-x_i)$ is the collected intensity, i.e. the number of detected photons, $n_i$,  $I_0$ is a proportionality factor that describes how the brightness of the emitter depends on the illumination laser power, and $b$ is the proportion of signal arising from the background. It is instructive to express $b$ as multiplied by $I_0$ since the background, which arises from auto- or out-of-focus fluorescence background or, in case of probing with a local minimum, from an imperfect zero in the PSF, will scale with illumination laser power. 

Let us investigate the precision with which we can localize a fluorophore from a single measurement, provided we know the brightness of the fluorophore $I_0$. For simplicity, we place the PSF at $x_i=0$ and neglect background ($b=0$). Then, the number of photons we detect is $N=I_0 f(x_0)$. We can determine the fluorophore position as $x_0= f^{-1}(N/I_0)$. In a linear approximation, the error of the position $\delta x_0$ is proportional to the error  $\delta N$ in measuring the photons $N$:

\begin{align} \label{eq:xx}
   \delta N=\frac{\partial I}{\partial x_0} \delta x_0=I_0 f'(x_0) \delta x_0,
\end{align}
where $f'(x_0)=\partial f / \partial x |_{x_0}$ is the derivative of $f$ along $x$, i.e., the gradient of the excitation PSF, evaluated at $x_0$.

Because the detection of photons is a Poisson process, $\delta N =\sqrt{N}$. We can substitute $I_0=N/f(x_0)$ and solve for $\delta x_0$:
\begin{align} \label{eq:first_order_precision}
  \delta x_0= \left|\frac{1}{\sqrt{N}}\frac{f(x_0)}{f'(x_0)}\right|,
\end{align}
where we have taken the absolute value since localization precision is always positive. Thus, in a first approximation, the position error scales inversely with $\sqrt{N}$ and is proportional to the ratio of magnitude $f(x_0)$ and gradient $f'(x_0)$ of the excitation PSF at the position of the emitter. For realistic measurements, the fluorophore brightness is typically not known, and we need to perform additional measurements.

To examine the localization precision in the case when we make sufficient measurements to capture fluorophore brightness, position, and background, we derive the localization precision using Fisher information and the Cram\'{e}r-Rao (lower) bound (CRB). 

The Fisher information is usually expressed as a matrix where each element is calculated as
\begin{align} \label{eq:fisher_standard}
    \mathcal{I}_{l,m}(x_0) = \sum_{i=1}^K \frac{1}{I(x_0-x_i)}\frac{\partial I(x_0-x_i)}{\partial\theta_l}\frac{\partial I(x_0-x_i)}{\partial\theta_m},
\end{align}
where $\theta \in \{I_0, x_0, b\}$. We call the number of unknown variables $M$, which in this case is $3$. In this paper, we slightly modify Eq. \ref{eq:fisher_standard} by following the framework outlined in Masullo et al. \cite{masullo_common_2022}. We assume pure Poisson noise and call $N$ the total number of photons detected over $K$ exposures with the excitation PSF at positions $x_i$. The probability of collecting $n_i$ photons during one of these measurements $i \in K$ is
\begin{align} \label{eq:prob}
    p_i(x_0) = \frac{I(x_0-x_i)}{\sum_{j=1}^K I(x_0-x_j)}.
\end{align}
We can now compute the elements of the Fisher information matrix as 
\begin{align}
    \mathcal{I}_{l,m}(x_0) = N\sum_{i=1}^K \frac{1}{p_i(x_0)}\frac{\partial p_i(x_0)}{\partial\theta_l}\frac{\partial p_i(x_0)}{\partial\theta_m},
\end{align}
where we have made the substitution $I(x_0-x_i) = Np_i(x_0)$ where
$N = \sum_{i=1}^K n_i$. This substitution eliminates $I_0$ from the Fisher information matrix (and wraps it in $N$), allowing us to compute the matrix for $M$ unknowns using $M-1$ variables \cite{eilers_minflux_2018}, $\theta \in \{x_0, b\}$. Even with this substitution, we still require $K \ge M$ exposures to compute $M$ unknown variables.

\subsection{In the case of known background}

Suppose we have previously measured the background offset $b$ in our experiment, or we can extract it from our measurements in a post-processing analysis. In this case, $I(x_0-x_i)$ as expressed in Eq. \ref{eq:Igen} has only two unknowns: $I_0$ and $x_0$. To estimate these unknowns from the measured number of photons $n_i$, $K$ must be $\geq 2$. The Fisher information matrix in this case is a single entry,
\begin{align}
    \mathcal{I}(x_0) = N\sum_{i=1}^K\frac{1}{p_i}
        \left(\frac{\partial p_i}{\partial x_0}\right)^2,
\end{align}
where $N = \sum_{i=1}^K n_i$ is the total number of photons detected over $K$ exposures and $p_i$ is as defined in Eq. \ref{eq:prob}.

The theoretically best achievable localization precision can be estimated as the square root of the CRB,  $\Sigma_\text{CRB}$, which is the inverse of the Fisher information matrix, $\mathcal{I}(x_0)$. The localization precision in the case of known background for $K=2$ is then
\begin{align} \label{eq:2pointcrb_known}
    \sigma_{x_0} = \sigma_\text{CRB} = \sqrt{\Sigma_\text{CRB}} = \sqrt{\mathcal{I}(x_0)^{-1}} = \sqrt{\frac{1}{N\left(\frac{1}{p_1}\left(\frac{\partial p_1}{\partial x_0}\right)^2 + \frac{1}{p_2}\left(\frac{\partial p_2}{\partial x_0}\right)^2\right)}},
\end{align}
and for $K=3$ it is
\begin{align} \label{eq:3pointcrb_known}
    \sigma_{x_0} = \sqrt{\frac{1}{N\left(\frac{1}{p_1}\left(\frac{\partial p_1}{\partial x_0}\right)^2 + \frac{1}{p_2}\left(\frac{\partial p_2}{\partial x_0}\right)^2 + \frac{1}{p_3}\left(\frac{\partial p_3}{\partial x_0}\right)^2\right)}}.
\end{align}

\subsubsection{Equivalence with signal-to-background ratio formulation}

In \cite{balzarotti_nanometer_2017, eilers_minflux_2018, masullo_common_2022}, and other places, the derivation for MINFLUX localization precision expresses the intensity response of an emitter as
\begin{align} \label{eq:Igen_nobg}
    I_\text{nobg}(x_0-x_i) = I_0f(x_0-x_i),
\end{align}
where $I_0$, $f(x)$, $x_i$, and $x_0$ are as in Eq. \ref{eq:Igen}. The background is included as a signal-to-background ratio
\begin{align}
    \label{eq:sbr}
    \text{SBR}(x_0) = \frac{\sum_{j=1}^K I_\text{nobg}(x_0-x_j)}{\sum_{j=1}^K b_j},
\end{align}
in $p_i(x_0)$, rather than in $I(x_0-x_i)$. By setting $b_j = b_0$ $\forall j$, it is possible to express $p_i(x_0)$ as
\begin{align} \label{eq:prob_sbr}
    p_i(x_0) = \frac{\text{SBR}(x_0)}{\text{SBR}(x_0)+1}\frac{I_\text{nobg}(x_0-x_i)}{\sum_{j=1}^K I_\text{nobg}(x_0-x_j)} + \frac{1}{\text{SBR}(x_0)+1}\frac{1}{K}.
\end{align}
We can recover $p_i(x_0)$ in the form of Eq. \ref{eq:prob} by substituting the definition of $\text{SBR}$ back in to Eq. \ref{eq:prob_sbr} and using the trivial substitution $b_0 = I_0b$, where $I_0$ is as defined in Eq. \ref{eq:Igen_nobg}. Then,
\begin{align}
    p_i(x_0) &= \frac{\frac{\sum_{j=1}^K I_\text{nobg}(x_0-x_j)}{KI_0b}}{\frac{\sum_{j=1}^K I_\text{nobg}(x_0-x_j)}{KI_0b}+1}\frac{I_\text{nobg}(x_0-x_i)}{\sum_{j=1}^K I_\text{nobg}(x_0-x_j)} + \frac{1}{\frac{\sum_{j=1}^K I_\text{nobg}(x_0-x_j)}{KI_0b}+1}\frac{1}{K} \\
    &= \frac{1}{\frac{\sum_{j=1}^K I_\text{nobg}(x_0-x_j)}{KI_0b}+1}\left(\frac{I_\text{nobg}(x_0-x_i)}{KI_0b} + \frac{1}{K}\right) \\
    &= \frac{1}{\frac{\sum_{j=1}^K I_\text{nobg}(x_0-x_j)}{KI_0b}+\frac{KI_0b}{KI_0b}}\left(\frac{I_\text{nobg}(x_0-x_i)}{KI_0b} + \frac{I_0b}{KI_0b}\right) \\
    &= \frac{I_\text{nobg}(x_0-x_i)+I_0b}{\sum_{j=1}^K I_\text{nobg}(x_0-x_j) + I_0b},
\end{align}
which is equivalent to Eq. \ref{eq:prob}. In \cite{balzarotti_nanometer_2017, eilers_minflux_2018, masullo_common_2022}, the derivative of the SBR is not considered in the Fisher information matrix. Thus, the SBR formulation and the known background formulation are equivalent.

\subsection{In the case of unknown background}

Suppose we have not previously measured our background offset. We want to understand how background influences our emitter localization precision, so we now include it in $\mathcal{I}(x_0)$: 
\begin{align}
    \mathcal{I}(x_0) = N\sum_{i=1}^K\frac{1}{p_i}
    \begin{bmatrix}
        \left(\frac{\partial p_i}{\partial x_0}\right)^2 & \frac{\partial p_i}{\partial x_0}\frac{\partial p_i}{\partial b} \\
        \frac{\partial p_i}{\partial b}\frac{\partial p_i}{\partial x_0} & \left(\frac{\partial p_i}{\partial b}\right)^2
    \end{bmatrix}.
\end{align}
In this case, $I(x_0-x_i)$ as defined in Eq. \ref{eq:Igen} has three unknowns: $I_0$, $x_0$, and $b$. To estimate these unknowns from the measured number of photons $n_i$, $K$ must be $\geq 3$. For $K=3$,
\begin{align}
    \mathcal{I}(x_0) = N
    \begin{bmatrix}
        \frac{1}{p_1}\left(\frac{\partial p_1}{\partial x_0}\right)^2 + \frac{1}{p_2}\left(\frac{\partial p_2}{\partial x_0}\right)^2 + \frac{1}{p_3}\left(\frac{\partial p_3}{\partial x_0}\right)^2 & 
        \frac{1}{p_1}\left(\frac{\partial p_1}{\partial x_0}\frac{\partial p_1}{\partial b}\right) + \frac{1}{p_2}\left(\frac{\partial p_2}{\partial x_0}\frac{\partial p_2}{\partial b}\right) + \frac{1}{p_3}\left(\frac{\partial p_3}{\partial x_0}\frac{\partial p_3}{\partial b}\right) \\
        \frac{1}{p_1}\left(\frac{\partial p_1}{\partial b}\frac{\partial p_1}{\partial x_0}\right) + \frac{1}{p_2}\left(\frac{\partial p_2}{\partial b}\frac{\partial p_2}{\partial x_0}\right) + \frac{1}{p_3}\left(\frac{\partial p_3}{\partial b}\frac{\partial p_3}{\partial x_0}\right) & 
        \frac{1}{p_1}\left(\frac{\partial p_1}{\partial b}\right)^2 + \frac{1}{p_2}\left(\frac{\partial p_2}{\partial b}\right)^2 + \frac{1}{p_3}\left(\frac{\partial p_3}{\partial b}\right)^2,
    \end{bmatrix}
\end{align}
and $\Sigma_\text{CRB}$ is
\begin{align}
    \Sigma_\text{CRB} = \frac{1}{cN}
    \begin{bmatrix}
        \frac{1}{p_1}\left(\frac{\partial p_1}{\partial b}\right)^2 + \frac{1}{p_2}\left(\frac{\partial p_2}{\partial b}\right)^2 + \frac{1}{p_3}\left(\frac{\partial p_3}{\partial b}\right)^2 & -\frac{1}{p_1}\left(\frac{\partial p_1}{\partial x_0}\frac{\partial p_1}{\partial b}\right) - \frac{1}{p_2}\left(\frac{\partial p_2}{\partial x_0}\frac{\partial p_2}{\partial b}\right) - \frac{1}{p_3}\left(\frac{\partial p_3}{\partial x_0}\frac{\partial p_3}{\partial b}\right) \\
        -\frac{1}{p_1}\left(\frac{\partial p_1}{\partial b}\frac{\partial p_1}{\partial x_0}\right) - \frac{1}{p_2}\left(\frac{\partial p_2}{\partial b}\frac{\partial p_2}{\partial x_0}\right) -   \frac{1}{p_3}\left(\frac{\partial p_3}{\partial b}\frac{\partial p_3}{\partial x_0}\right) &
        \frac{1}{p_1}\left(\frac{\partial p_1}{\partial x_0}\right)^2 + \frac{1}{p_2}\left(\frac{\partial p_2}{\partial x_0}\right)^2 + \frac{1}{p_3}\left(\frac{\partial p_3}{\partial x_0}\right)^2,
    \end{bmatrix}
\end{align}
where
\begin{align}
    c = \left(\frac{1}{p_2}\frac{1}{p_3}\left(\frac{\partial p_3}{\partial x_0}\frac{\partial p_2}{\partial b}-\frac{\partial p_2}{\partial x_0}\frac{\partial p_3}{\partial b}\right)^2+\frac{1}{p_1}\left(\frac{1}{p_2}\left(\frac{\partial p_2}{\partial x_0}\frac{\partial p_1}{\partial b} - \frac{\partial p_1}{\partial x_0}\frac{\partial p_2}{\partial b}\right)^2 + \frac{1}{p_3}\left(\frac{\partial p_3}{\partial x_0}\frac{\partial p_1}{\partial b}-\frac{\partial p_1}{\partial x_0}\frac{\partial p_3}{\partial b}\right)^2\right)\right).
\end{align}
The localization precision $\sigma_{x_0}$ is given by the standard deviation $\sigma_\text{CRB,i,j} = \sqrt{\Sigma_\text{CRB,i,j}}$ for the upper left element,
\begin{align} \label{eq:3pointcrb}
     \sigma_{x_0} = \sigma_\text{CRB,1,1}= \sqrt{\frac{\frac{1}{p_1}\left(\frac{\partial p_1}{\partial b}\right)^2 + \frac{1}{p_2}\left(\frac{\partial p_2}{\partial b}\right)^2 + \frac{1}{p_3}\left(\frac{\partial p_3}{\partial b}\right)^2}{N\left(\frac{1}{p_2}\frac{1}{p_3}\left(\frac{\partial p_3}{\partial x_0}\frac{\partial p_2}{\partial b}-\frac{\partial p_2}{\partial x_0}\frac{\partial p_3}{\partial b}\right)^2+\frac{1}{p_1}\left(\frac{1}{p_2}\left(\frac{\partial p_2}{\partial x_0}\frac{\partial p_1}{\partial b} - \frac{\partial p_1}{\partial x_0}\frac{\partial p_2}{\partial b}\right)^2 + \frac{1}{p_3}\left(\frac{\partial p_3}{\partial x_0}\frac{\partial p_1}{\partial b}-\frac{\partial p_1}{\partial x_0}\frac{\partial p_3}{\partial b}\right)^2\right)\right)}}.
\end{align}
We can see that, as in the simpler cases, the accuracy of the emitter position estimate is dependent on the derivatives of $p_i(x_0)$, and thus dependent on the derivatives of $I(x_0-x_i)$ with respect to $x_0$, $\partial I(x_0-x_i)/\partial x_0$, which is proportional to the excitation PSF gradient at the emitter position.

\section{Localization precision of emitters in 1D}

Now armed with analytical expressions, we can explicitly compute the localization precision of emitters in 1D for specific PSFs.

\subsection{Minimum}

The quadratic equation is a good approximation of the center of the donut PSF commonly used in MINFLUX \cite{balzarotti_nanometer_2017}. We define a 1D quadratic PSF with background as
\begin{align} \label{eq:quad}
    I_\text{quad,1D}(x_0-x_i) = I_0\left(\frac{(x_0-x_i)^2}{2\sigma_q^2} + b\right),
\end{align}
where $I_0$, $x_i$, $x_0$, and $b$ are as defined in Eq. \ref{eq:Igen} and $\sigma_q$ parameterizes the steepness of the PSF. It is important to note that $\sigma_q$ cannot be smaller than what is allowed by the diffraction limit of the system optics ($\sim$\qty{250}{\nano\meter}).

For $b=0$, the localization precision of the simple single-point measurement (Eq. \ref{eq:first_order_precision}) is
\begin{align}
    \label{eq:first_order_precision_quad}
    \delta x_0 =\frac{x_0}{2\sqrt{N}}.
\end{align}

\subsubsection{Minimum with known background}

Suppose we know the background, e.g. through measurement of the average response $I_\text{quad,1D}(x_0-x_i)$ in empty space. In this case, we can substitute Eq. \ref{eq:quad} evaluated at $x_i=-L/2$, and $L/2$, where $L \in \mathcal{R}$ describes the displacement between the two measurements, into Eqs. \ref{eq:prob} and \ref{eq:2pointcrb_known} to get
\begin{align}
    \label{eq:2pointminimanonnormal_known}
    \sigma_{x_0} = \frac{1}{4L\sqrt{N}} \sqrt{\frac{\left(8 b \sigma _q^2+L^2+4 x_0^2\right){}^2 \left(16 b \sigma _q^2 \left(4 b \sigma _q^2+L^2+4 x_0^2\right)+\left(L^2-4 x_0^2\right)^2\right)}{ \left(8 b \sigma _q^2+L^2-4 x_0^2\right){}^2}}.
\end{align}
For $b=0$, this simplifies to
\begin{align} \label{eq:2pointminimanonnormal_known_b0}
    \sigma_{x_0}(b=0) = \frac{L}{4\sqrt{N}}\left[1+\left(\frac{x_0}{L/2}\right)^2\right],
\end{align}
which is equivalent to the corresponding Eq. S22b found in the supplement of \cite{balzarotti_nanometer_2017}. If we further let $x_0 = 0$, we get 
\begin{align} \label{eq:2pointminimanonnormal_known_x0_b0}
    \sigma_{x_0}(x_0=0,b=0) = \frac{L}{4\sqrt{N}}.
\end{align}
Note that the single-measurement approximation Eq. \ref{eq:first_order_precision_quad} measured at $x_0 = L/2$ is equivalent to Eq. \ref{eq:2pointminimanonnormal_known_x0_b0}. 

In the case Eq. \ref{eq:quad} is evaluated at three points, $x=-L/2$, $0$, and $L/2$, and substituted into into Eqs. \ref{eq:prob} and \ref{eq:3pointcrb_known}, we get
\begin{align} \label{eq:3pointminimanonnormal_known}
    \sigma_{x_0} = \frac{1}{4L\sqrt{N}}\sqrt{\frac{2\left(2 b \sigma _q^2+x_0^2\right) \left(12 b \sigma _q^2+L^2+6 x_0^2\right){}^2 \left(16 b \sigma _q^2 \left(4 b \sigma _q^2+L^2+4 x_0^2\right)+\left(L^2-4 x_0^2\right)^2\right)}{\left(4 b \sigma _q^2 \left(4 b \sigma _q^2 \left(24 b \sigma _q^2+5 L^2-12 x_0^2\right)+L^4+6 L^2 x_0^2-24 x_0^4\right)+3 x_0^2 \left(L^2-4 x_0^2\right)^2\right)}}.
\end{align}
At $x_0=0$, Eq. \ref{eq:3pointminimanonnormal_known} simplifies to
\begin{align} \label{eq:3pointminimax0_known}
    \sigma_{x_0}(x_0=0) = \frac{1}{4L\sqrt{N}} \sqrt{(L^2+8b\sigma_q^2)(L^2+12b\sigma^2)},
\end{align}
and in the case of no background ($b=0$), Eq. \ref{eq:3pointminimax0_known} simplifies to
\begin{align}
    \sigma_{x_0}(x_0=0,b=0) = \frac{L}{4\sqrt{N}},
\end{align}
which is equivalent to the value of the two-point case at the same position and background (Eq. \ref{eq:2pointminimanonnormal_known_x0_b0}). 

\subsubsection{Minimum with an unknown background}

Suppose we have not measured our background signal and we measure at three points, $x_i=-L/2$, $0$, and $L/2$. We plug $I_\text{quad,1D}(x_0-x_i)$ into equations \ref{eq:prob} and \ref{eq:3pointcrb} to get
\begin{align} \label{eq:3pointminimanonnormal}
    \sigma_{x_0} = \frac{1}{4L^2 \sqrt{N}} \sqrt{\left(12 b \sigma_q^2+L^2+6 x_0^2\right) \left(8 b \left(L^2+48 x_0^2\right) \sigma_q^2+L^4-12 L^2 x_0^2+192 x_0^4\right)}.
\end{align}

\begin{SCfigure}[0.6][h!]
\includegraphics[width=0.5\textwidth]{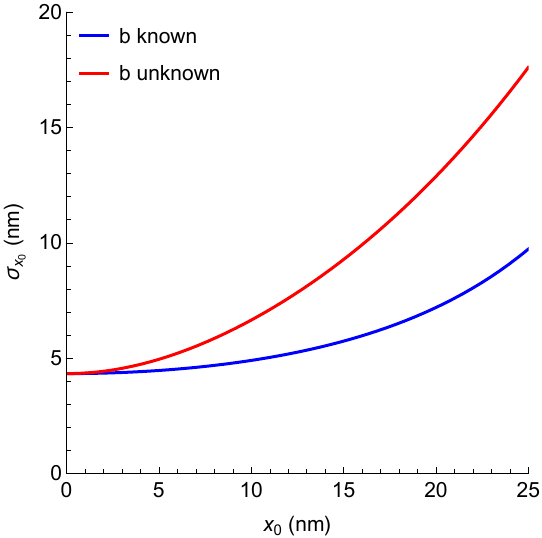}
\caption{Comparison of the localization precision of an emitter probed at three points with a quadratic minimum in the cases of assumed unknown (Eq. \ref{eq:3pointminimanonnormal}) and known (Eq. \ref{eq:3pointminimanonnormal_known}) background, evaluated over emitter position $x_0$  for $L=$ \qty{50}{\nano\meter}, $\sigma_q =$ \qty{250}{\nano\meter}, $b=0.01$ and $N=100$ photons. \label{fig:known_vs_unknown}}
\end{SCfigure}

At $x_0=0$, Eq. \ref{eq:3pointminimanonnormal} simplifies to \ref{eq:3pointminimax0_known}, suggesting that the known and unknown background cases are equivalent. However, away from $x_0 = 0$, the localization precision in the cases of assumed unknown (Eq. \ref{eq:3pointminimanonnormal}) and assumed known (Eq. \ref{eq:3pointminimanonnormal_known}) background differ significantly, as shown in Figure \ref{fig:known_vs_unknown}. Knowing the background (i.e. having more information) results in a better localization precision.

For the example shown in Figure \ref{fig:known_vs_unknown} where the background value is known, it is possible to convert directly between $b$ and $\text{SBR}$ via substitution:
\begin{align*}
    \text{SBR}(x_0=0) &= \frac{\sum_{j=1}^K I(x_j)}{Kb} \\
    &= \frac{I(-50/2) + I(0) + I(50/2)}{(3)(0.01)} \\
    &\approx 1.33.
\end{align*}

\subsubsection{Localization precision in the presence of different background sources}

The fluorescence background $b$ (Eq. \ref{eq:Igen}) in a MINFLUX experiment can come from 1) an imperfect zero in the excitation PSF or 2) auto-fluorescence of the sample or optics or out-of-focus signal from other fluorophores. 

\begin{SCfigure}[0.6][h!]
\includegraphics[width=0.5\textwidth]{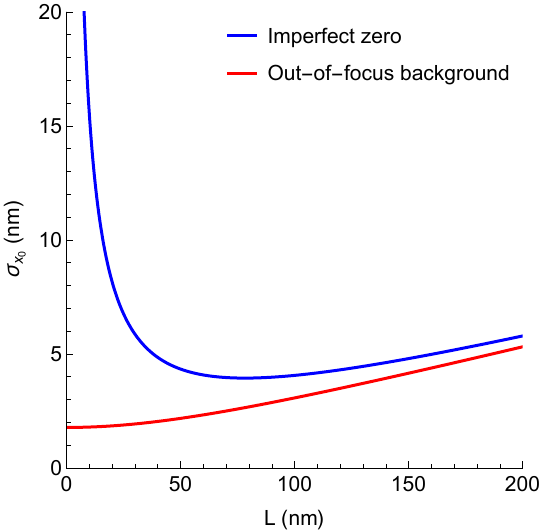}
\caption{Comparison of $\sigma_{x_0}$ vs. $L$ at $x_0=0$ for an emitter probed with a quadratic minimum at three points and background coming from either an imperfect zero (Eq. \ref{eq:imperfect_zero}) or from 
 autofluorescence or out-of-focus emitters (Eq. \ref{eq:out_of_focus}). Here, $b=0.01$, $\sigma_q =$ \qty{250}{\nano\meter} and $N_s=100$. \label{fig:background}}
\end{SCfigure}

Let's consider the behavior of background at $x_0=0$. In the case of an imperfect zero, all photons come from the photon budget of the target fluorophore, $N_s$. Thus, we substitute $N = N_s$, the number of signal photons, in Eq. \ref{eq:3pointminimax0_known} to get
\begin{align}\label{eq:imperfect_zero}
    \sigma_{x_0}(x_0=0) = \frac{1}{4L\sqrt{N_s}}\sqrt{(L^2+8b\sigma_q^2)(L^2+12b\sigma_q^2)}.
\end{align}
In the case of auto-fluorescence or out-of-focus signal, the background photons, $N_b$, do not originate from the target fluorophore. We substitute $N=N_s+N_b$ into Eq. \ref{eq:3pointminimax0_known}. We then make the observation that for the $K=3$ case, when we know the background, $N_b=3 b N_s/(I(-L/2)+I(0)+I(L/2))$. In this case,
\begin{align}
    b_{x_0=0} = \frac{L^2N_b}{12N_s\sigma_q^2}.
\end{align}
Subsequently, plugging this in to our $N = N_s + N_b$--substituted version of Eq. \ref{eq:3pointminimax0_known} we get
\begin{align} \label{eq:out_of_focus}
    \sigma_{x_0}(x_0=0) = \frac{1}{4\sqrt{N_s}}\sqrt{L^2+8b\sigma_q^2},
\end{align}
in the case we have autofluorescence or out-of-focus background. We plot a comparison of this and and the imperfect zero case in Figure \ref{fig:background}.
There is a general trend that as $L$ decreases, the localization precision improves, as expected. However, in the case of background arising from an imperfect zero, too small of an $L$ is dominated by background. We can also see that shrinking $L\to 0$ does not result in perfect localization precision in the case of autofluorescence or out-of-focus background, and there is a minimum uncertainty generated by the background.

\subsection{Maximum}
Next, we look at 1D localization with a maximum, corresponding to using confocal-style microscopes. The diffraction-limited, excitation, Airy-function-shaped PSF can be approximated closely by a Gaussian function: 
\begin{align}\label{eq:gauss}
    I(x_0-x_i)_\text{Gauss,1D} = I_0\left(e^{-\frac{(x_0-x_i)^2}{2\sigma_q^2}} + b\right),
\end{align}
where $I_0$, $x_i$, $x_0$, and $b$ are as defined in Eq. \ref{eq:Igen} and, as in Eq. \ref{eq:quad}, the diffraction limit sets a lower bound on the steepness of the PSF, $\sigma_q$. 

\begin{SCfigure}[0.6][h!]
\includegraphics[width=0.5\textwidth]{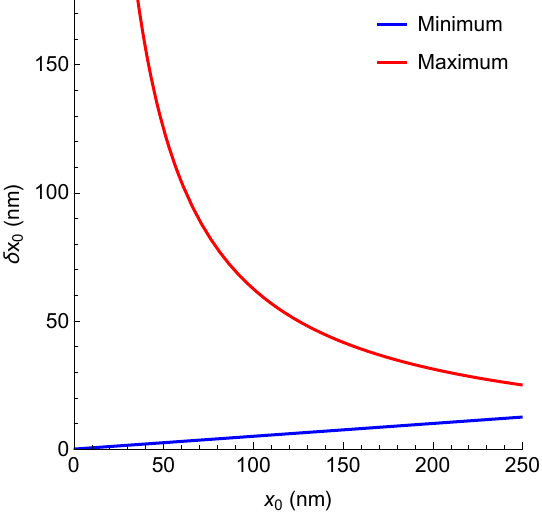}
\caption{Comparison of first-order emitter localization precision (Eq. \ref{eq:first_order_precision}) arising from minimum (Eq. \ref{eq:quad}) and maximum (Eq. \ref{eq:gauss}) excitation PSFs evaluated at $\sigma_q =$ \qty{250}{\nano\meter}, $N=100$, and $x_0=0$ over a field of view of length $\sigma_q$. \label{fig:maxima_vs_minima_derivative_normalized}}
\end{SCfigure}

For $b=0$, the localization precision of a simple single-point measurement with a Gaussian maximum (Eq. \ref{eq:first_order_precision}) is
\begin{align}
    \label{eq:first_order_precision_gauss}
    \delta x_0 = \frac{\sigma_q^2}{x_0\sqrt{N}}.
\end{align}
We can use this simple estimate to get an intuition for how a maximum (Eq. \ref{eq:gauss}) will perform as compared to a minimum (Eq. \ref{eq:quad}). We can see in Figure \ref{fig:maxima_vs_minima_derivative_normalized} that the precision in the minimum case (Eq. \ref{eq:first_order_precision_quad}) approaches $0$ 
for $x_0=0$, while the precision for the maximum (Eq. \ref{eq:first_order_precision_gauss}) improves as it moves further from the emitter position. This tells us 1) the minimum is more sensitive to position than the maximum in the standard measurement range $L<\sigma_q$ and 2) an emitter can be probed at a shorter distance with a minimum excitation PSF, reducing sparsity requirements.

\subsubsection{Maximum with a known background}

\begin{SCfigure}[0.6][h!]
\includegraphics[width=0.5\textwidth]{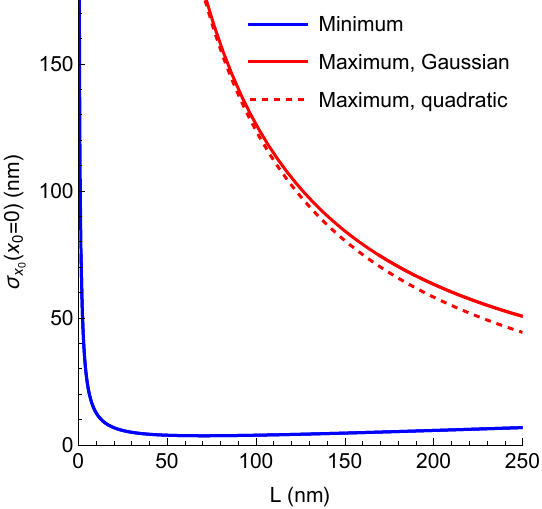}
\caption{Comparison of localization precision of maximum and minimum probing of an emitter with two point measurements in 1D and with known background. The plot shows $\sigma_{x_0}$ from Eqs. \ref{eq:2pointgauss} (maximum, Gaussian), \ref{eq:2pointnegquad} (maximum, quadratic) and \ref{eq:2pointminimanonnormal_known} (minimum) evaluated at $x_0=0$ vs. separation distance $L$ for $\sigma_q =$ \qty{250}{\nano\meter}, $b=0.01$, and $N=100$ photons. Note that for the optimal choice of $L$, $L << \sigma_q$, the minimum outperforms the maximum.\label{fig:minima_vs_maxima}}
\end{SCfigure}

Suppose we again measure at two points, $x_i=-L/2$ and $L/2$. When the separation distance $L < \sigma_q$, the maximum of the Gaussian can be approximated as a quadratic equation,
\begin{align} \label{eq:quad_neg}
    I_\text{negquad,1D}(x_0-x_i) = I_0\left(1-\frac{(x_0-x_i)^2}{2\sigma_q^2} + b\right),
\end{align}
and so
\begin{align} \label{eq:2pointnegquad}
    \sigma_{x_0} = \frac{1}{4L\sqrt{N}} \sqrt{\frac{\left(-8 (b+1) \sigma _q^2+L^2+4 x_0^2\right){}^2 \left(-16 (b+1) \left(L^2+4 x_0^2\right) \sigma _q^2+64 (b+1)^2 \sigma _q^4+\left(L^2-4 x_0^2\right)^2\right)}{ \left(-8 (b+1) \sigma _q^2+L^2-4 x_0^2\right){}^2}}.
\end{align}
This has worse localization precision than the corresponding minimum when $L < \sigma_q$, as shown in Figure \ref{fig:minima_vs_maxima}.
At $x_0=0$ and with no background ($b=0$), Eq. \ref{eq:2pointnegquad} simplifies to
\begin{align} 
    \sigma_{x_0} (x_0=0,b=0) = \frac{1}{4L\sqrt{N}} \left|L^2-8 \sigma _q^2\right|.
\end{align}

In the range of $L < \sigma_q$, the illumination profile will be near its peak and $N$ will be much higher than in the case of equivalent probing with a minimum. That is, for a fixed $L$, when $L < \sigma_q$, the emitter will experience a much greater intensity in the maximum case and emit more photons, resulting in a poorer photon efficiency than in the minimum case. 

\begin{SCfigure}[0.6][h!]
\centering
\includegraphics[width=0.5\textwidth]{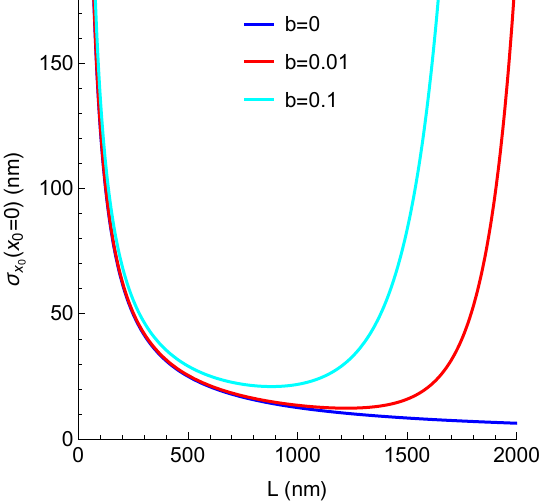}
\caption{Localization precision for probing an emitter at two points with a Gaussian maximum. The plot shows $\sigma_{x_0}$ from Eq. \ref{eq:2pointgauss} vs. separation distance $L$ for $x_0=0$, $\sigma_q =$ \qty{250}{\nano\meter} and $N=100$ photons at 3 different background levels.\label{fig:2pointgauss}}
\end{SCfigure}

For $L > \sigma_q$, it may be possible to illuminate the emitter with the less intense parts of two Gaussians. In this case, the quadratic approximation does not apply. Using the Gaussian PSF (Eq. \ref{eq:gauss}), we obtain a lengthy expression, which for $x_0=0$ simplifies to
\begin{align} \label{eq:2pointgauss}
    \sigma_{x_0}(x_0=0) = \frac{\sigma_q}{\sqrt{N}}\frac{2\sigma_q}{L} \left(b e^{\frac{L^2}{8 \sigma _q^2}}+1\right).
\end{align}
We can see in Figure \ref{fig:minima_vs_maxima} that the position uncertainty of the Gaussian maximum initially decreases with increasing displacement $L$. However, as shown in Figure \ref{fig:2pointgauss}, the gradient and magnitude of the Gaussian tails eventually diminish to a point where, in the presence of background, the localization precision gets worse again. This is influenced by the Gaussian width $\sigma_q$, which is fundamentally limited by diffraction, and by the strength of the background. We can see in Figure \ref{fig:minima_vs_maxima} that for the optimal choice of $L$, $L << \sigma_q$, the minimum outperforms the maximum.

For the no background case, the Gaussian PSF CRB simplifies to
\begin{align} 
    \label{eq:2pointgauss_nobg}
    \sigma_{x_0}(b=0) = \frac{\sigma_q}{\sqrt{N}}\frac{2\sigma_q}{L} \cosh \left(\frac{L x_0}{2 \sigma _q^2}\right),
\end{align}
which is equivalent to the corresponding Eq. S23b in the supplement of \cite{balzarotti_nanometer_2017} in the event that the definition of the Gaussians are made to be in agreement; this means exchanging $\frac{4\log(2)}{fwhm^2}$ in Eq. S19 of the supplement of \cite{balzarotti_nanometer_2017} for $\frac{1}{2\sigma_q^2}$ to match our Eq. \ref{eq:gauss} (change from calculating via the full width at half maximum to the standard deviation).

Without background ($b=0$), Eq. \ref{eq:2pointgauss} simplifies to 
\begin{align} 
    \sigma_{x_0}(x_0=0,b=0) = \frac{2\sigma_q^2}{L\sqrt{N}},
\end{align}
which is equivalent to the corresponding Eq. S24b in the supplement of \cite{balzarotti_nanometer_2017} and equivalent to Eq. \ref{eq:first_order_precision_gauss} evaluated at $x_0 = L/2$.

\section{Localization precision of emitters in 2D}

The CRB in 2D is almost identical to what was derived in Section \ref{sec:1dcrb}, but we now swap $y_0$ for $b$ in the Fisher information matrix. We also now compute the average localization precision, $\sigma_\text{CRB} = \sqrt{\frac{1}{2}\text{Tr}\left(\Sigma_\text{CRB}\right)}$, to examine the contributions along both the $x$ and $y$ directions. We will leave $b$ in the definition of $I(x_0-x_i, y_0-y_i)$ as a known constant. 

\subsection{2D MINFLUX with explicit, known background}

We want to show that our notation is equivalent to published results \cite{balzarotti_nanometer_2017, eilers_minflux_2018, masullo_common_2022} in 2D as well as 1D. 2D MINFLUX localization precision is often calculated for four measurements of an emitter at $(x_i, y_i) = (0,0), (-\frac{L}{4}, \frac{\sqrt{3}L}{4}), (-\frac{L}{4},-\frac{\sqrt{3}L}{4}), (\frac{L}{2}, 0)$ using a donut,
\begin{align} \label{eq:2d_donut}
    I_\text{donut}(x_0-x_i,y_0-y_i) = I_0\left(e\frac{(x_0-x_i)^2+(y_0-y_i)^2}{2\sigma_q^2}\exp\left(-\frac{(x_0-x_i)^2+(y_0-y_i)^2}{2\sigma_q^2}\right) + b\right),
\end{align}
where $I_0$ and $b$ are as defined in Eq. \ref{eq:Igen}, the diffraction limit sets a lower bound on the steepness of the PSF, $\sigma_q$, $(x_0,y_0)$ is the position of the emitter and $(x_i, y_i)$ is the position of the excitation PSF. Computing the CRB for Eq. \ref{eq:2d_donut} at these points and evaluating it at $(x_0, y_0) = (0,0)$ yields
\begin{align}
    \label{eq:donut2d}
    \sigma_{x_0}(x_0=0,y_0=0) = \frac{\sigma_q}{\sqrt{N}}\frac{\sqrt{8}\sigma_q}{L} \sqrt{\frac{L^2\left(eL^2+8b\exp\left(\frac{L^2}{8\sigma_q^2}\right)\sigma_q^2\right)\left(3eL^2+32b\exp\left(\frac{L^2}{8\sigma_q^2}\right)\sigma_q^2\right)}{3e^2(L^3-8L\sigma_q^2)^2}}.
\end{align}

In \cite{balzarotti_nanometer_2017, eilers_minflux_2018, masullo_common_2022}, there is no background term in the definition of the donut (Eq. \ref{eq:2d_donut}). Substituting $b = (I_\text{donut}(x_0, y_0) + I_\text{donut}(x_0+\frac{L}{4}, y_0-\frac{\sqrt{3}L}{4}) + I_\text{donut}(x_0+\frac{L}{4}, y_0+\frac{\sqrt{3}L}{4}) + I_\text{donut}(x_0-\frac{L}{2}, y_0))/(4I_0\text{SBR}(x_0))$, using the definition of SBR (Eq. \ref{eq:sbr}), yields
\begin{align}
    \label{eq:donut2d_x0y0}
    \sigma_{x_0}(x_0=0,y_0=0) = \frac{\sigma_q}{\sqrt{N}}\frac{\sqrt{8}\sigma_q}{L}\frac{L^2}{8\sigma_q^2-L^2}\sqrt{\left(\frac{1}{\text{SBR}(x_0)}+1\right) \left(\frac{3}{4 \text{SBR}(x_0)}+1\right)}
\end{align}
If, as for Eq. \ref{eq:2pointgauss_nobg}, we assume a Gaussian full width at half maximum, we can make the substitution $\frac{1}{2\sigma_q^2} = \frac{4\log(2)}{fwhm^2}$. With this substitution, Eq. \ref{eq:donut2d_x0y0} is equivalent to Eq. S31 in the supplement of \cite{balzarotti_nanometer_2017}. Evaluating Eq. \ref{eq:donut2d} at $b=0$ yields
\begin{align}
    \sigma_{x_0}(x_0=0,y_0=0) = \frac{\sigma_q}{\sqrt{N}} \frac{\sqrt{8}\sigma_q}{L} \frac{L^2}{ \left(8 \sigma _q^2-L^2\right)},
\end{align}
which is equivalent to its corresponding equation S27 in the supplement of \cite{balzarotti_nanometer_2017}, again in the case we set $\frac{1}{2\sigma_q^2} = \frac{4\log(2)}{fwhm^2}$.

\subsection{Orbital tracking}

In 2D MINFLUX, we measure about an emitter with a donut-shaped minimum. In orbital tracking, instead a Gaussian PSF is used \cite{enderlein_tracking_2000}. In fluorophore tracking, we expect the emitter to move and update our orbit about the emitter to be centered on its new position $(x_i, y_i)$ at each time point $i$. For ease of derivation, we will consider here the case where the excitation PSF orbits a stationary emitter. In this case, the 2D Gaussian PSF is expressed as
\begin{align}\label{eq:gauss2d}
    I_\text{Gauss,2D}(x_0-x_i,y_0-y_i) = I_0\left(\exp\left(-\frac{(x_0-x_i)^2+(y_0-y_i)^2}{2\sigma_q^2}\right) + b\right),
\end{align}
where $I_0$, is the amplitude of the illumination, ($x_i$, $y_i$) is the coordinate of the excitation PSF, ($x_0$, $y_0$) is the coordinate of the emitter, $\sigma_q$ parameterizes the steepness of the PSF, and $b$ is the proportion of signal arising from the background. 

For most orbital tracing, the movement is continuous and the number of measurements $K$ is large. For ease of computation, let us consider three (or four, or six, or twelve) points evenly spaced on a circle with radius $L/2$, centered on the emitter location $(x_0,y_0)$. Following the framework established in Section \ref{sec:1dcrb}, using $y_0$ instead of $b$ in the Fisher information matrix, and setting $(x_0,y_0)=(0,0)$, we get 
\begin{align}\label{eq:ot}
    \sigma_{x_0,y_0}(x_0=0, y_0=0) = \frac{\sigma_q}{\sqrt{N}} \frac{\sqrt{8}\sigma_q}{L}  \left(b e^{\frac{L^2}{8 \sigma_q^2}}+1\right).
\end{align}
Interestingly, because we achieve the same result for $K=3,4,6$ and $12$,this implies that the CRB at $(x_0, y_0) = (0,0)$, is independent of the choice of $K$ when $\sigma_{q}$ is the same along $x$ and $y$. 

\begin{SCfigure}[0.6][h!]
\centering
\includegraphics[width=0.5\textwidth]{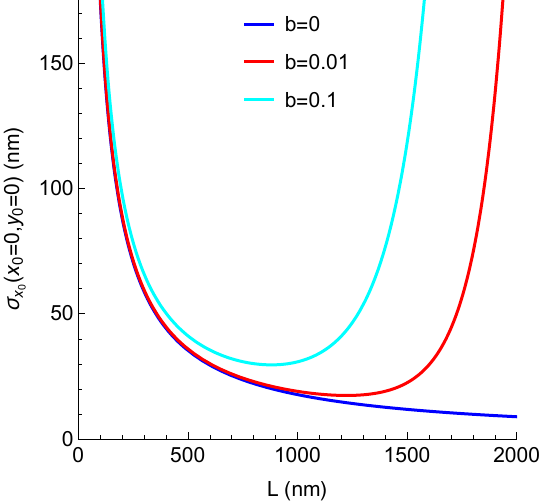}
\caption{Localization precision for probing an emitter with an orbital tracking scheme. The plot shows $\sigma_{x_0}$ from Eq. \ref{eq:ot} vs. separation distance $L$ for $x_0=0$, $\sigma_q =$ \qty{250}{\nano\meter}, and $N=100$ photons at 3 different background levels.\label{fig:ot}}
\end{SCfigure}

Without background, Eq. \ref{eq:ot} suggests that the localization precision can be improved by letting $L\to\infty$. However, any background leads to a deterioration of the localization precision with large $L$ (Figure \ref{fig:ot}), leading to an optimal value of $L\approx5\sigma_q$ \cite{masullo_common_2022, enderlein_tracking_2000}. This phenomenon is shown in Figure \ref{fig:ot} and is in agreement with the behavior of the two Gaussian measurement in the 1D case (Eq. \ref{eq:2pointgauss}).

\section*{Conclusion}

In this paper, we proposed an alternative formulation to the standard signal-to-background approach for background in MINFLUX measurements that allows us to investigate the influence of unknown background on localization precision. We found that this formulation agrees with previous work when the background is known. 

\section*{Acknowledgements}

The authors thank Dr. Sheng Liu, Nikolay Sergeev, Dr. Takahiro Deguchi, Dr. Lukas Scheiderer, and Dr. Francesco Reina for helpful discussions.


\printbibliography

@article{eilers_minflux_2018,
	title = {{MINFLUX} monitors rapid molecular jumps with superior spatiotemporal resolution},
	volume = {115},
	issn = {0027-8424, 1091-6490},
	url = {https://pnas.org/doi/full/10.1073/pnas.1801672115},
	doi = {10.1073/pnas.1801672115},
	language = {en},
	number = {24},
	journal = {Proceedings of the National Academy of Sciences},
	author = {Eilers, Yvan and Ta, Haisen and Gwosch, Klaus C. and Balzarotti, Francisco and Hell, Stefan W.},
	month = jun,
	year = {2018},
	pages = {6117--6122},
}

@article{masullo_common_2022,
	title = {A common framework for single-molecule localization using sequential structured illumination},
	volume = {2},
	issn = {26670747},
	url = {https://linkinghub.elsevier.com/retrieve/pii/S2667074721000367},
	doi = {10.1016/j.bpr.2021.100036},
	language = {en},
	number = {1},
	journal = {Biophysical Reports},
	author = {Masullo, Luciano A. and Lopez, Lucía F. and Stefani, Fernando D.},
	month = mar,
	year = {2022},
	pages = {100036},
}

@article{balzarotti_nanometer_2017,
	title = {Nanometer resolution imaging and tracking of fluorescent molecules with minimal photon fluxes},
	volume = {355},
	issn = {0036-8075, 1095-9203},
	url = {https://www.science.org/doi/10.1126/science.aak9913},
	doi = {10.1126/science.aak9913},
	language = {en},
	number = {6325},
	journal = {Science},
	author = {Balzarotti, Francisco and Eilers, Yvan and Gwosch, Klaus C. and Gynnå, Arvid H. and Westphal, Volker and Stefani, Fernando D. and Elf, Johan and Hell, Stefan W.},
	month = feb,
	year = {2017},
	pages = {606--612},
}

@article{enderlein_tracking_2000,
	title = {Tracking of fluorescent molecules diffusing within membranes},
	volume = {71},
	issn = {0946-2171, 1432-0649},
	url = {http://link.springer.com/10.1007/s003400000409},
	doi = {10.1007/s003400000409},
	language = {en},
	number = {5},
	journal = {Applied Physics B},
	author = {Enderlein, J.},
	month = nov,
	year = {2000},
	pages = {773--777},
}

\end{document}